\documentclass[conference]{IEEEtran}
\IEEEoverridecommandlockouts

\usepackage{subcaption}
\usepackage{tikz-cd}

\usepackage{mathtools}
\usepackage{mwe}

\usepackage{minted}

\usepackage{cite}
\usepackage{amsmath,amssymb,amsfonts}
\usepackage{algorithmic}
\usepackage{graphicx}
\usepackage{textcomp}
\usepackage{xcolor}
\def\BibTeX{{\rm B\kern-.05em{\sc i\kern-.025em b}\kern-.08em
    T\kern-.1667em\lower.7ex\hbox{E}\kern-.125emX}}

\usepackage{hyperref}
\usepackage{cleveref}

\begin{document}

\title{Parametrized Exact Hardware-Software Partitioning}

\author{
\IEEEauthorblockN{Cameron Ibrahim\IEEEauthorrefmark{1},
S M Ferdous\IEEEauthorrefmark{3},
Erdal Mutlu\IEEEauthorrefmark{2},
Ilya Safro\IEEEauthorrefmark{1} and
Mahantesh Halappanavar\IEEEauthorrefmark{2}}
\IEEEauthorblockA{\IEEEauthorrefmark{1}Department of Computer Science\\
University of Delaware, Newark, DE, United States\\
Email: \{cibrahim, isafro\}@udel.edu}
\IEEEauthorblockA{\IEEEauthorrefmark{2}Pacific Northwest National Lab, Richland, WA, United States\\
Email: \{erdal.mutlu, Mahantesh.Halappanavar\}@pnnl.gov }
\IEEEauthorblockA{\IEEEauthorrefmark{3}UNC Charlotte, Charlotte, NC, United States\\
Email: sferdou1@charlotte.edu}
}

\maketitle

\begin{abstract}
When optimizing computing architecture for specific computationally intensive tasks, such as evaluating or training a neural network, it is important to identify what computational subtasks will offer the greatest decrease in cost (e.g., wall clock time or energy usage). This problem is known as Hardware-Software (HS) Partitioning, and it has a variety of formulations, many of which are NP-Hard.  
In this paper, we will define a family of HS Partitioning formulations which admit an exact fixed parameter tractable algorithm based on the directed pathwidth of the given task graph and show that this family of problems contains multiple existing formulations such as makespan minimization. Finally, we examine task graphs with small directed pathwidth that arise in real world applications and show that our algorithm can provide a speed up of up to 200x over a comparable linear programming approach utilizing the Gurobi ILP Library.\\
\noindent{\bf Reproducibility:} 
Our source code and data are available at \url{https://github.com/cameton/HPEC26_HSPart}.
\end{abstract}

\begin{IEEEkeywords}
Graph Decomposition, Hardware-software codesign, Fixed parameter tractability
\end{IEEEkeywords}

\section{Introduction}

The ability to build and utilize custom hardware accelerators for challenging computational tasks is becoming increasingly prevalent in the modern computing landscape. Examples include combinatorial optimization \cite{liu2022leveraging}, machine learning \cite{sze2017hardware,liu2022towards}, simulation \cite{xiao2019survey}, and  information partitioning \cite{liu2022partitioning} to mention just a few. Whether designing a custom chiplet or deciding when to utilize a Tensor Processing Unit, Hardware-Software (HS) Partitioning is a crucial stage in process wherein we attempt to identify which parts of the computational task benefit from hardware acceleration \cite{vahid_whatisHS}.

A computational task graph is a directed graph that represents a complex computation as a set of tasks (vertices) and their data or control dependencies (edges), making explicit the order and parallelism inherent in the workflow. Broadly, a given formulation of HS Partitioning will attempt to find a subset of vertices in the tasks graph which minimizes some metric for cost given a budget \cite{arato_hs}. There are two types of vertices that correspond to software and hardware computation, respectively. Generally, the ``Software'' vertices will represent subtasks which will be executed as instructions on a a microprocessor or collection of microprocessors, while the ``Hardware'' vertices represent tasks which will be executed on some sort of \emph{custom} chiplet (e.g., FPGA or GPU \cite{vahid_whatisHS}). Common metrics for cost include wall clock time, energy usage, chip size, and monetary costs. Similarly, bounds can be placed on any of these metrics.

There are a number of relevant NP-Hard formulations of the HS Partitioning problem depending on what costs need to be minimized and what resource constraints are present in your setting. This means it can be intractable to get an exact solution for some generic problem instances while using these formulations.  In this paper, we will focus primarily on the minimization of the span of the computational task graph, that is the amount of time the task graph would take to execute with an unbounded number of parallel processes \cite{papadimitriou1988towards, critical_path_thesis, arato_hs}. 

\textbf{Our contribution:} In this paper, we identify a class of HS Partitioning problems which are actually fixed parameter tractable to solve; that is, the cost required to compute a solution is determined primarily by the \emph{directed pathwidth} of the computational task graph, a graph parameter which can be used to measure the complexity of a given input. To do so, we provide the following contributions.
\begin{itemize}
    \item In \cref{sec:dp}, we define a recurrence for the HS Partitioning problem based on a directed path decomposition of the task graph, enabling a dynamic programming algorithm that provides a speed up of up to 200x relative to an equivalent LP formulation.
    \item In \cref{sec:dpd}, we introduce a solver for finding directed path decompositions based on nested bisection utilizing a novel formulation for balanced vertex separators in directed acyclic graphs.
    \item In \cref{sec:experiments}, we examine of real world task graphs generated from popular neural network architectures using the SODA toolchain to demonstrate that task graphs with low directed pathwidths do arise naturally \cite{SODA, resnet18}. Neural networks such as RESNET have a pathlike structure, which is  reflected in the computational task graph.
\end{itemize}
Our results directly support the broader goals of automated design exploration 
by providing a structure-aware method that improves the computational performance using HS partitioning. By revealing tractable regimes in realistic computational task graphs and demonstrating practical speedups, our approach enables more scalable optimization pipelines for emerging heterogeneous architectures.


\section{Background \& Notation}

For two integers $i$ and $j$ such that $i\leq j$, we use the notation \([i, j] \coloneqq \{i, i+1, \ldots, j\}\). Furthermore, we define \([j] \coloneqq [1, j]\) for $j\geq 1$. The set of all subsets of a given input set \(S\) is known as the power set \(\mathcal{P}(S).\) 

\subsection{Directed Path Decompositions}


A graph \(G = (V, E)\) consists of a set of vertices \(V\) alongside a set of edges \(E \subseteq \binom{V}{2}.\) A directed graph instead has edges \(E \subseteq V \times V.\) A graph is acyclic if there is no series of edges from which you can traverse a vertex to arrive back at itself. If a graph is directed and acyclic, it is known as a DAG. A separator is a set of vertices \(S \subseteq V\) such that there are disjoint sets \(A,B\) such that \(A \cup B = V \setminus S\) with no edges passing between \(A\) and \(B.\)

\begin{figure}[ht]
    \centering
    \begin{subfigure}{\linewidth}
        \centering
        \includegraphics[width=0.5\textwidth]{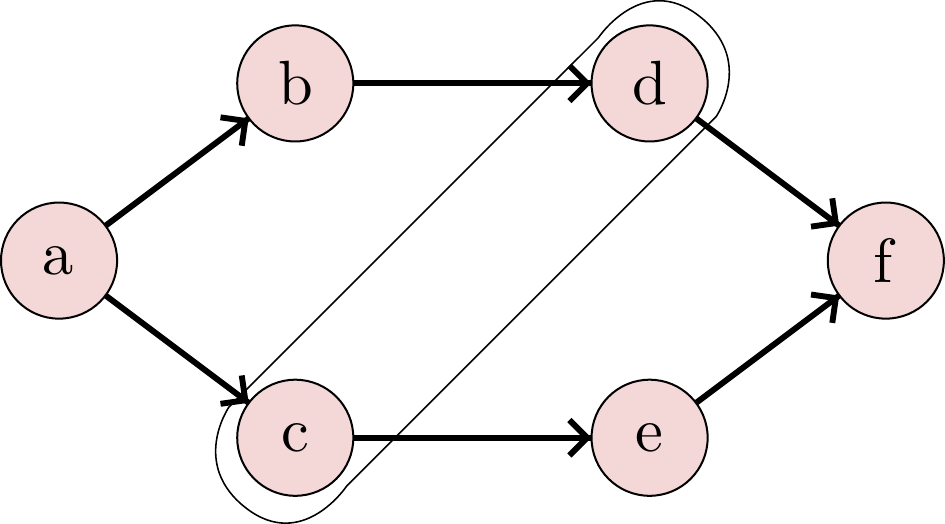}
        \caption{}
        \label{fig:example_task}
    \end{subfigure}\\
    \begin{subfigure}{\linewidth}
        \centering
        \begin{tikzcd}
        a \arrow[r] & ab \arrow[d]  & bcd \arrow[r] & \tikz{\node[draw, circle]{cd};} \arrow[d]  & def \arrow[r] & ef \arrow[d] \\
                    & abc \arrow[r] & bc \arrow[u]  & cde \arrow[r] & de \arrow[u]  & f           
        \end{tikzcd}
        \caption{}
        \label{fig:example_decomp}
    \end{subfigure}
    \caption{\cref{fig:example_task} is an example task graph \(G\), while \cref{fig:example_decomp} gives an example directed path decomposition of \(G.\) The separator \(\{c, d\}\) is circled in both figures.}
    \label{fig:path_decomp}
\end{figure}

For a graph \(G\) with \(n = \lvert V \rvert\) vertices, a \emph{path decomposition} on \(G\) is a family of \(N\) sets \(X_i \subseteq V\) for \(i \in [N]\) such that 
\begin{itemize}
    \item $\forall$ \(v \in V,\) \(v \in X_i \cap X_j\) implies \(v \in X_k\) for each \(k \in [i, j],\)
    \item $\forall$ \(uv \in E\) there exists an \(X_i\) such that \(u,v \in X_i\)
\end{itemize}
\cite{althaus2020optimal,bodlaender1996efficient}. Conceptually, a path decomposition is a collection of overlapping vertex separators forming a path structure, as shown in \cref{fig:path_decomp}.
The width of a path decomposition is \(\max_i \lvert X_i\rvert.\)

For this paper, we specifically consider path decompositions where any adjacent separators differ by only a single vertex, with the two separators at either end containing only one vertex; for any path decomposition, there is a path decomposition with this property that has the same width \cite{althaus2020optimal}. In this case, we have \(N = 2n.\)

Path decompositions are one of a number of structural graph decompositions that are key to the development of the field of fixed parameter tractable algorithms \cite{cygan2015parameterized}. These are algorithms for which there exists a parameter \(k\) of the input such that the computational complexity of the algorithm grows polynomially with \(n\) so long as \(k\) remains bounded.

One such family of algorithms utilizes a dynamic programming approach on a nice path decomposition in order to solve problems whose solution come in the form of a subsets of vertices. \(X_1\) is designated as the root, while all other nodes are categorized as follows:
\begin{itemize}
    \item if \(\lvert X_i\rvert < \lvert X_{i-1}\rvert\), \(X_i\) is a Forget node,
    \item if \(\lvert X_i\rvert > \lvert X_{i-1}\rvert\), \(X_i\) is an Introduce node.
\end{itemize}
A dynamic programming algorithm may then be constructed recursively on these separators, requiring an amount of memory exponential in the width of the path decomposition, but linear with the size of the graph \cite{althaus2020optimal}.

Recently, a directed variant of path decompositions has been introduced known as a directed path decomposition \cite{erde2017directedpathdecompositions}.  If \(G\) is a directed acyclic graph, a path decomposition \(X\) on \(G\) is a directed path decomposition if for every \(uv \in E,\) there exists \(i \leq j\) such that \(u\) appears in separator \(i\) and \(v\) appears in separator \(j\). An example of a directed path decomposition is given in \cref{fig:path_decomp}.

\subsection{HS Partitioning \& Related Work}

Hardware-Software (HS) Partitioning is a long studied problem in automated hardware design concerned with identifying which tasks in a computational graph should be run as instructions on a standard microprocessor and which tasks would benefit most from the construction of custom hardware to accelerate their execution \cite{vahid_whatisHS}. The former are collectively referred to as Software tasks, while the latter are referred to as Hardware tasks. Software and Hardware tasks are chosen to minimize some cost metric, such as execution time or power usage of the circuit. This same problem formulation may also be used to determine whether or not to use a hardware accelerator such as a GPU or TPU for a given task. There may also be additional constraints placed on the problem, such as setting a budget for how many tasks can be placed on hardware or limiting the total hardware area available. 

HS Partitioning has a wide array of possible formulations, many depending on a specific choice of application domain. What's more, similar to general graph partitioning \cite{buluc2016recent}, many constrained variants of HS Partitioning are NP-Hard, making exact solutions difficult if not impossible for large problem instances. As a result, a wide variety of heuristic methods for HS Partitioning have been introduced, such as Tabu Search\cite{hotpath_dp, search_survey, system_level}, Ant Colony Optimization \cite{ants}, and Genetic Algorithms \cite{genetic_algorithm, genetic_partitioning}.

In this paper, we are concerned primarily with \emph{exact} methods for solving HS Partitioning. Exact HS Partitioning approaches largely consist of methods which are solved using mixed integer linear programming \cite{arato_hs, ilp_accurate_high_level, integrate_constraints} and methods utilizing dynamic programming \cite{hotpath_dp, graph_reduce_hs}. Of these, the method presented by Shi et. al is most similar to the approach presented in this paper \cite{hotpath_dp}. There, the authors identify a single ``hot'' path in the input task graph to optimize. However, their dynamic programming approach operates only on graphs with pathwidth exactly 2. This is a significant limitation on where their approach can be applied, which we address by allowing graphs with pathwidth of any small \(k.\)

Another approach presented by Jiang et. al. aims to solve the scalability issues common with dynamic programming approaches by introducing a heuristic graph reduction method for compressing an input to a more tractable size \cite{graph_reduce_hs}. The main limitation of this approach is that its effectiveness depends strongly on the quality and quantity of reducible sub-graphs present in the task graph; applications with fewer tightly connected regions will benefit less from reduction. In addition, the greedy reduction   cannot guarantee optimality.

We focus our comparison in this paper on ILP formulations of HS partitioning. Furthermore, throughout this paper, we will work under the assumption that input computational graph is a directed graph with no cycles. 
In this paper, we focus on the minimization of makespan, which is the length of the critical path in the computational task graph \cite{critical_path_thesis}. This formulation aims to minimize the amount of time the task graph would take to execute with unbounded parallel resources.

We denote  \(C_{uv}\) to be the cost to communicate between tasks \(u\) and \(v\) when placed in different partitions, \(h_v\) to be the cost of a task \(v\) when placed in the hardware partition, \(s_v\) to be  the cost of task \(v\) when placed in the Software partition, and \(c_v\) to be the amount of area on the custom hardware that would be taken up by task \(v.\) For each vertex \(v,\) $x_v$ indicates the partition \(v\) is assigned to (e.g., 0 - software, 1 - hardware), while for each edge \(uv\), \(y_{uv}\) is 1 if tasks \(u\) and \(v\) are assigned to different partitions and zero otherwise. This yields the linear program which attempts to optimize \(\min_{x,y,t} \max_v t_v\) such that 
\begin{align*}
    &\begin{aligned}
    \forall uv \in E,\quad & t_u + C_{uv} x_{uv} + h_vx_v + s_v(1-x_v) \leq t_v\\
    & y_{uv} \leq x_u + x_v\\
    & y_{uv} \leq 2 - (x_u + x_v)\\
    & y_{uv} \in \{0, 1\}
    \end{aligned}\\
    &\forall v \in V, x_v \in \{0, 1\}, t_v \geq 0.
\end{align*}
where the constraints  force the solution to form a proper partition. This formulation of HS Partitioning remains NP-Hard despite the lack of resource constraints \cite{papadimitriou1988towards}.

Execution of many machine learning models consist of a large sequential series of arithmetic operations which typically form a pathlike structure \cite{liu2017survey}. While more research is needed to examine the full breadth of relevant pathwidths that may appear in applications across industries, the low pathwidth subgraphs that appear in neural network architectures such as RESNET make a compelling case for the development of algorithms that take better advantage of this structure \cite{resnet18}.

\section{Methodology}

In this section, we introduce a recurrence for the HS partitioning problem based on directed path decompositions which captures multiple existing formulations and  develop a dynamic programming algorithm to exactly solve this recurrence. We then introduce a novel solver for finding a directed path decomposition based on recursive bisection. 

\subsection{Dynamic Programming Algorithm}\label{sec:dp}

The key observation we utilize in this paper is that for a given separator of the computational graph which separates any initial tasks from the final result, the cost to execute the computational graph can be neatly written as: (1)  the cost to execute everything up to the separator, (2) the cost to execute the tasks in the separator itself, and (3) the cost to execute everything after the separator. Given a sequential series of separators, we should then be able to write the cost of evaluating the computational graph recursively.

Because a directed path decomposition consists of a series of overlapping separators, we can then write the HS Partitioning cost as a recursion on these separators \(X_i\) with \(i \in [2n]\), starting from the final output of the computational graph \(X_{2n}\).  We refer to the proposed partition for a given separator as \(p\), the width of separator \(X_i\) as \(k_{i-1}\), to the currently used area as \(a.\) \(\Delta\) is a formulation specific update function which updates the cost of the previous separator based on the current solution. All other notation is shared with the LP formulation.
\begin{align*}
        \mathcal{L}(p, a) &\coloneqq \min_{p^\prime \colon X_{i-1} \to \{0, 1\}}  \begin{cases}
        \Delta\left(L, p\right) & p\vert_{X_{i-1}} = p^\prime\\
        \mathcal{L}(p^\prime, a) & p = p^\prime \vert_{X_i}\\
        \infty & T < a
    \end{cases}
\end{align*}
where \(L = \mathcal{L}(p^\prime, a \oplus  c_v(p))\) and \(\{v\} = X_i\setminus X_{i-1}.\)

Notably, we don't need to choose just a single scalar for the output of each function. In particular, we can define an ordering on vectors \(X_i \to \mathbb{R}\) for each \(X_i.\) 


Here we define a recursion which seeks to compute the minimal makespan of a given computational task graph. When working with makespan, we assume that there is no limit on hardware area, i.e. \(T = \infty.\) For each \(v,u \in V\) define \(h_v, s_v,C_{uv}\) representing hardware cost, software cost, and communication cost for a given partitioning of a given separator. These are defined such that \(h_v(p) = 0\) if \(p(v) = 1,\) \(s_v(p) = 0\) if \(p(v) = 2,\) and \(C_{uv}(p) = 0\) if \(p(v) = p(u)\) or \(uv \notin E.\) Let us define 
\begin{align*}
    \Delta(L, p)(u) &\coloneqq \begin{cases}
        \max\{ L(z) + C_{zv}(p) \mid zv \in E\}  & u = v\\
        \qquad + h_v(p) + s_v(p)\\
        L(u) & u \neq v
    \end{cases}
\end{align*}
In this way, the time cost of a given task is the maximum amount of time to compute and transfer all of it's inputs to the same device as \(v,\) plus the time to run \(v.\)

Because there is no limit on hardware area, 
constructing a table to memorize results can be done in \(O(nf(k))\) time, where \(f(k) = O(k2^k)\) and \(k\) is the directed pathwidth of the decomposition used. 





\begin{figure*}
    \centering
    \includegraphics[width=0.7\linewidth]{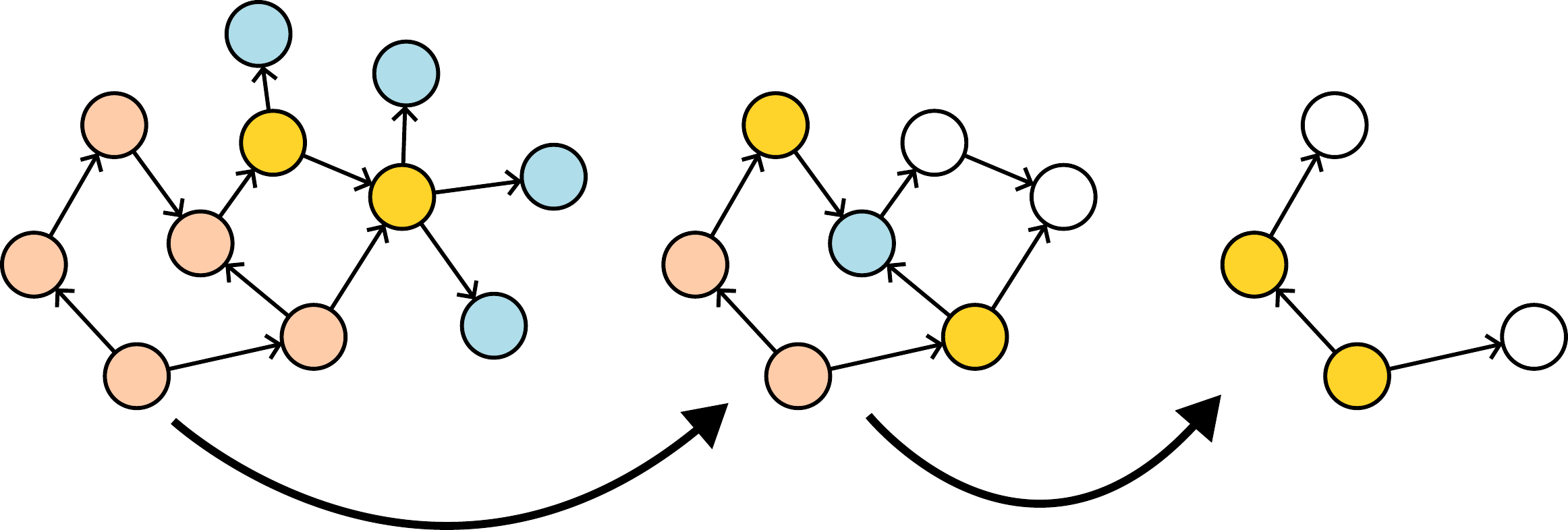}
    \caption{A depiction of the directed path decomposition constructing algorithm. The branch of recursion following the topologically earlier partitions is shown. For each iteration, we provide the subgraph induced by the interior of the candidate (colored vertices), as well as the boundary of the candidate (empty vertices). At each step of the iteration, a directed separator is found (yellow vertices), which partitions the candidate set into a set of vertices which precede the separator (pink) and a set of vertices which follow the separator (blue). The boundary of the next candidate is a subset of the boundary of the current candidate and the separator for the current candidate. When a separator can't be found, the remaining interior is returned as a solution for that candidate.
    }    
    \label{fig:recursion}
\end{figure*}

\subsection{Directed Path Decompositions}\label{sec:dpd}

To find good directed path decomposition solvers, we leverage a strategy similar to that used by \cite{flowcutter, sep_approx}. This approach centers around recursively finding minimal balanced separators in a graph, and using these to construct the individual separators in the directed path decomposition. This process is visualized in \cref{fig:recursion}. We find that we are able to take advantage of the DAG structure of the input graphs to find a fairly simple LP formulation of the balanced directed separator problem.

\subsubsection{Balanced Directed Separators}\label{sec:sep}

To find minimal balanced separators in a DAG, we present a modified version of the linear programming formulation utilized in \cite{althoby2020exact}. As a separator is a subset of vertices whose removal spits a graph into two disjoint parts, one can identify the size of the separator from the total number of vertices in those individual parts, yielding the objective
\begin{align*}
    \min_{x,y} \lvert V \rvert - \lVert x\rVert_1  - \lVert y \rVert_1
\end{align*}
where \(x,y \in \{0, 1\}^{\lvert V \rvert}\).
The key modification we make over a standard formulation for balanced separators is to require one partition to come topologically before the vertex separator and the other partition to come after it. 

Formally, we achieve this by requiring that every incoming edge to the separator has its tail in \(x\) and every outgoing edge for the separator has its tail in \(y.\) Equivalently, we can say that if a vertex has its tail in \(y,\) its head must also be in \(y\) and vice versa for \(x.\) This yields the constraint
\begin{align*}
    \forall uv \in E : x_u \geq x_v,\qquad y_u \leq y_v.
\end{align*}
Additionally, we forbid an edge from having its tail in \(x\) and its head in \(y\), meaning that no edges pass directly between \(x\) and \(y\)
\begin{align*}
    \forall uv \in E,\quad & x_u + y_v \leq 1.
\end{align*}
The final constraint enforcing correctness of the separator requires that no vertex can be in both \(x\) and \(y\)
\begin{align*}
    \forall v \in V,\quad & x_v + y_v \leq 1
\end{align*}
To prevent trivial solutions, we require each partition to have at least 1 vertex and to obey an upper bound on size known as a balance constraint
\begin{align*}
    1 \leq \lVert x\rVert_1 \leq \beta \qquad 1 \leq \lVert y\rVert_1 \leq \beta
\end{align*}
With this formulation, we can then utilize a linear programming solver such as Gurobi to efficiently find a balanced separator for the input. However,  for very large instances a more scalable heuristic will be required. For example, vertex separator problems have  been addressed using scalable multilevel  methods  \cite{hager2018multilevel}.

\subsubsection{Decomposition Solver}

In order to construct our directed path decomposition for a DAG \(G\), we utilize a heuristic strategy based on the tree decomposition solver Flowcutter \cite{flowcutter}. This strategy works via recursive bisection, iteratively splitting the input graph to construct a hierarchy of separators.

At each step in the iteration, we have a candidate separator \(B \cup I,\) where \(B\) is known as the boundary of the separator and \(I\) as the interior of the separator. For the first iteration \(B = \emptyset\) and \(I = V.\) We utilize our balanced vertex separator formulation from the previous section on the induced subgraph \(G\vert_I\), which yields partitions \(I_1\) and \(I_2\) as well as a separator \(s\). If \(S = I\), or if no valid separator can be found, \([B \cup I]\) is returned as the solution.

\(I_1\) and \(I_2\) will serve as the interiors of future candidates. Their boundaries \(B_1, B_2 \subseteq B\cup S\) are the vertices in the current boundary which contain a vertex adjacent to a vertex in \(I_1\) or \(I_2\) respectively. We then apply the algorithm recursively on \(B_1 \cup I_1\) to get a path decomposition \(\ell,\) and on \(B_2 \cup I_2\) to get a path decomposition \(r.\) Then, \(\ell\) is concatenated with \([B \cup S]\) and \(r\) and returned as a path decomposition. 

\section{Experiments}\label{sec:experiments}

To evaluate our approach, we first compare our dynamic programming algorithm with the linear programming formulation of Makespan HS Partitioning on a collection of synthetic graphs with bounded pathwidth. We follow this by an examination of the task graphs of a number of well known Neural Network architectures, and show that they contain significant subgraphs with a very small directed pathwidth. For the linear programming experiments, we solve the problem both with the Gurobi and GLPK solvers with a 10 minute time limit in order to evaluate how well the state of the art for linear programming compares to a more accessible alternative. All experiments are performed with an AMD Ryzen 5 7640U. 

 \begin{figure*}[ht]
    \centering
    \begin{subfigure}{0.8\textwidth}
        \centering
        \includegraphics[width=1\linewidth]{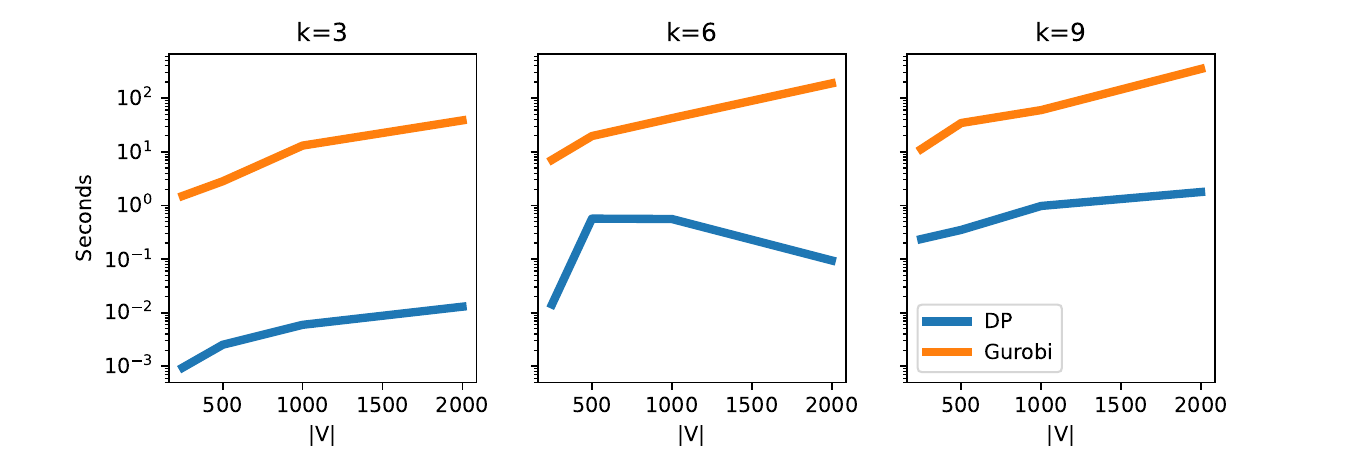}
        \caption{}
        \label{subfig:series}
    \end{subfigure}\\
    \begin{subfigure}{0.8\textwidth}
        \centering
        \includegraphics[width=1\linewidth]{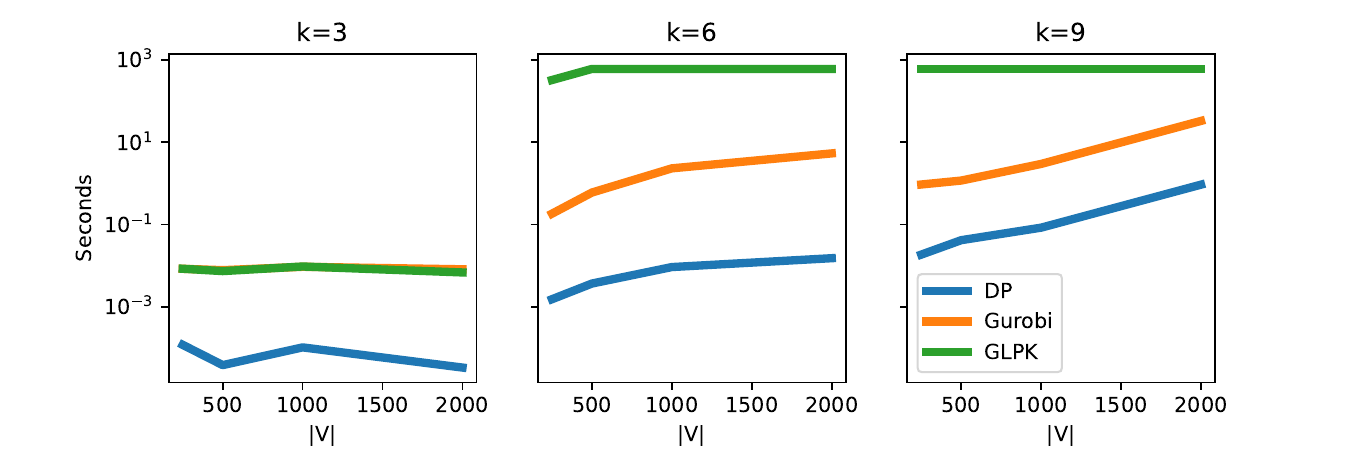}
        \caption{}\label{subfig:mart}
    \end{subfigure}
    \caption{A comparison of running time in seconds on a log scale between the dynamic programming (DP) and linear programming (LP) approaches between two sets of synthetic graphs with bounded pathwidth. \Cref{subfig:series}  compares dynamic programming and Gurobi for a collection of series-parallel graphs, while \cref{subfig:mart} compares dynamic programming, Gurobi, and GLPK for martingale graphs. The plots corresponding to martingale graphs have pathwidth \(k\), while each plot for the series-parallel graphs corresponds to a pathwidth \(k+1.\) We see that the dynamic programming approach is better on every instance, with an improvement of up to 200x for the largest of the series-parallel graphs. A time limit of 600s was set for both algorithms.}    \label{fig:time_comparison}
\end{figure*}

\subsection{Synthetic Graphs}

 The goal of these synthetic data experiments is to demonstrate the scaling behavior of the dynamic programming algorithm. We aim to show that so long as the directed pathwidth of the input remains small, the dynamic programming algorithm can outperform competitive solvers even for large inputs. To do so, we consider the two different types of synthetic graphs. Each graph tested is given uniformly random weights to represent the communication cost between nodes, the hardware time, and the software time. Software time weights are given 2x the size of the corresponding hardware time estimates.

 \subsubsection{Series-Parallel Graphs}

 In this section, we examine chains of series-parallel graphs where each graph in the chain is made up of \(k\) length \(3\) paths with a shared start and end point \cite{eppstein1992parallel}. In  \cref{fig:time_comparison}, we see the comparison between dynamic programming and Gurobi on a collection of series-parallel graphs with  \(k \in \{3,6,9\}\) parallel paths and \(\ell \in \{250, 500, 1000, 2000\}\) layers. GLPK timed out on some of the smallest instance, so GLPK is excluded from this plot.

We see that while the dynamic programming approach maintains a negligible cost throughout, the time required by the linear programming begins to be considerable for the larger graphs in the group. For the largest graph, the dynamic programming approach is almost 200\% faster then Gurobi for this instance. For emphasis, we also find the optimal solution for a task graph with 2000 layers using the dynamic programming approach, a size considerably larger than could be comfortably handled by the ILP.

\subsubsection{Martingale Graphs}

Next we consider a collection of graphs that we generate by directly forming their directed path decomposition by adding and removing nodes in such a way that the size of the bag at any given point in the decomposition is expressible as a martingale. We refer to this construction as a martingale graph. In \cref{fig:time_comparison}, we compare the results of our dynamic programming algorithm with those achieved by Gurobi and GLPK. GLPK failed to solve a majority of instances and repeatedly hit the set time limit. On the other hand, Gurobi performed fairly well on these instances, taking only 36s on the largest input compared to the 300+s taken on the largest input of the chains of theta graphs. The dynamic programming algorithm on the other hand had sub second times for nearly all instances. 
\emph{The dynamic programming beats the linear programming approach on almost every instance. Notably, it is able to take only a few seconds to compute results for problem instances which caused the linear programming approach to exceed the 10 minutes timeout period.}





\subsection{Neural Networks}

We've mentioned previously that the layer construction of many neural network architectures forms something that is fairly pathlike, consisting of chains of matrix operations in sequence. 
We see the core tasks of the RES NET Neural Network, generated using the SODA toolchain \cite{SODA}. This small subgraph represents a series of convolution, multiplication, and subtraction operations and results in a computational task graph with a directed pathwidth of only 5. While this particular subgraph is trivially easy for both the linear programming and dynamic programming approaches to solve, we can see the benefits of this approach when we chain a large number of small pathwidth subgraphs together.

In this section, we aim to show this formally by finding a directed path decomposition for a number of common neural network architectures. These are represented as tasks graphs, which have been generated by the SODA toolchain \cite{SODA}.

In \cref{tab:path_width}, we see the directed pathwidths for a collection of neural network architectures with and without ``torch.const'' operations. With constant operations, some of the graphs such as Mobile Net and RESNET have a directed pathwidth too high to realistically apply any sort of fixed parameter tractable algorithm. However, we also see that without these constant operations, the highest directed pathwidth of any of these architectures is actually 6, something very near to being a path. As these constant operations do not contribute meaningfully to the makespan of the taskgraph, this means that every one of these architectures is amenable to our dynamic programming approach in this context. 

\begin{figure}
    \centering   \includegraphics[width=1.0\linewidth]{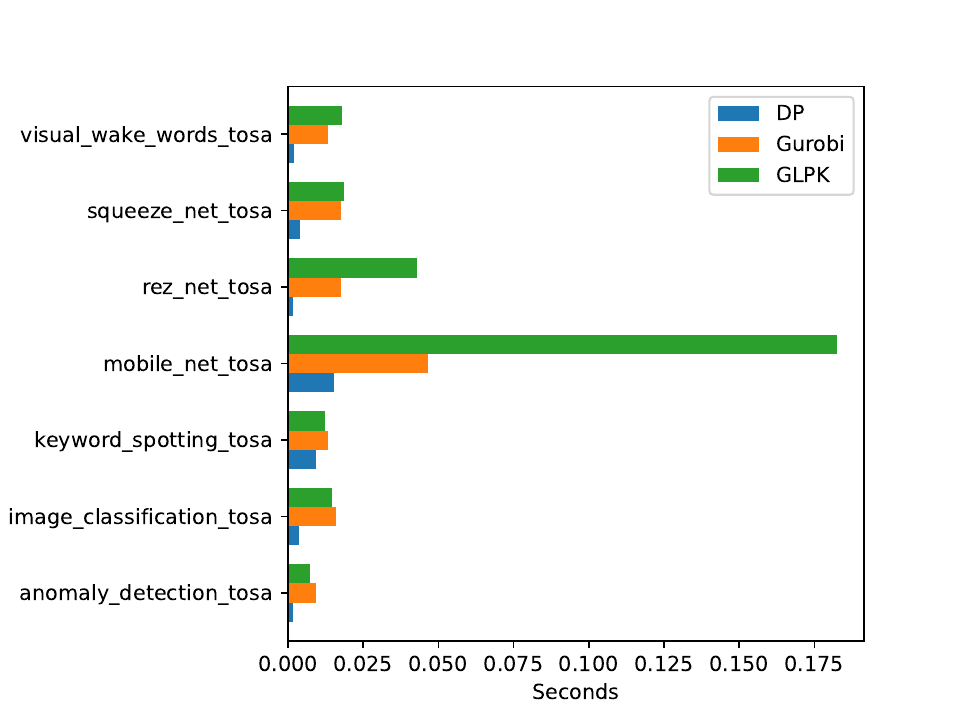}
    \caption{Comparison of dynamic programming (DP), Gurobi, and GLPK on a collection of graphs generated from real neural network architectures. Because of their sparsity and low directed pathwidth, these graphs were fairly straightforward to solve for all solvers.}
    \label{fig:real_data}
\end{figure}

We explore this further in \cref{fig:real_data}, where we compare the dynamic programming approach to LP approach for this batch of real world graphs. As these inputs are fairly simple, all tested solvers were able to find solutions in sub-second times. Together with the results from the synthetic data with much dense and more complicated graphs with bounded directed pathwidth, there is clearly a path forward with regards to targeting more complicated architectures with this approach.

\begin{table}[ht]
\centering
\begin{tabular}{ccccc}
                            & \multicolumn{2}{c}{Input}               & \multicolumn{2}{c}{No Const}            \\
                            & \(\lvert V\rvert\) & \(\lvert k\rvert\) & \(\lvert V\rvert\) & \(\lvert k\rvert\) \\ \hline
anomaly\_detection\_tosa    & 64                 & 5                  & 40                 & 4                  \\
image\_classification\_tosa & 121                & 8                  & 87                 & 6                  \\
keyword\_spotting\_tosa     & 119                & 8                  & 81                 & 5                  \\
mobile\_net\_tosa           & 826                & 117                & 591                & 5                  \\
rez\_net\_tosa              & 358                & 49                 & 251                & 5                  \\
squeeze\_net\_tosa          & 175                & 28                 & 121                & 4                  \\
visual\_wake\_words\_tosa   & 207                & 8                  & 134                & 5                 
\end{tabular}
\label{tab:path_width}
\caption{Directed path widths of a collection of neural network architectures. The size of each graph is given alongside the width, both for the original input graph and for the graph resulting from removing ``torch.const'' operations.}
\end{table}


\section{Conclusion}
We demonstrated that solving classical formulations of the HS Partitioning task become efficiently solvable when parameterized by the directed pathwidth of the underlying task graph. By leveraging this structural property, we developed a dynamic programming algorithm that delivers more than 200$\times$ speedup over conventional linear programming approaches. We further introduced a random graph generator tailored to produce low–pathwidth instances and showed that real-world task graphs, such as those derived from RESNET, naturally contain substantial low pathwidth subgraphs. This shows that exploiting graph structure can significantly accelerate partitioning decisions, enabling more scalable and effective hardware–software co-optimization for heterogeneous  systems.

Future work includes several interesting direction. First, we suggest extending this framework to richer cost models that incorporate additional architectural constraints, such as communication overhead or heterogeneous memory structures. 
Additionally, exploring realistic graph generators that learn from existing computationally inefficient HS partitioning scenarios (e.g., \cite{gutfraind2015multiscale,staudt2017generating}) could provide valuable benchmark families and guide the design of more robust algorithms. Such generators would allow systematic stress testing of partitioning strategies and help reveal structural bottlenecks present in real applications. 
Another promising direction is developing automated pathwidth-aware compilation pipelines that detect low–pathwidth regions in large task graphs and selectively apply fixed-parameter techniques to further accelerate real-world design flows.

\noindent{\bf Reproducibility:} 
Our source code and data are available at \url{https://github.com/cameton/HPEC26_HSPart}.

\section{Acknowledgments}

This work was supported in part by the U.S. Department of Energy, Office of Science, Office of Advanced Scientific Computing Research Competitive Portfolios program at Pacific Northwest National Laboratory (PNNL).  PNNL is a multi-program national laboratory operated for the U.S. Department of Energy (DOE) by Battelle Memorial Institute under Contract No. DE-AC05-76RL01830. 


\bibliographystyle{IEEEtran}
\bibliography{encode,ilya-biblio}

@article{vahid_whatisHS,
author = {Vahid, Frank},
title = {What is hardware/software partitioning?},
year = {2009},
issue_date = {June 2009},
publisher = {Association for Computing Machinery},
address = {New York, NY, USA},
volume = {39},
number = {6},
issn = {0163-5743},
url = {https://doi.org/10.1145/1862900.1862901},
doi = {10.1145/1862900.1862901},
journal = {SIGDA Newsl.},
month = jun,
pages = {1},
numpages = {1}
}

@article{althoby2020exact,
  title={Exact and heuristic methods for the vertex separator problem},
  author={Althoby, Haeder Y and Biha, Mohamed Didi and Sesbo{\"u}{\'e}, Andr{\'e}},
  journal={Computers \& Industrial Engineering},
  volume={139},
  pages={106135},
  year={2020},
  publisher={Elsevier}
}

@inproceedings{sep_approx,
author = {Feige, Uriel and Hajiaghayi, MohammadTaghi and Lee, James R.},
title = {Improved approximation algorithms for minimum-weight vertex separators},
year = {2005},
isbn = {1581139608},
publisher = {Association for Computing Machinery},
address = {New York, NY, USA},
url = {https://doi.org/10.1145/1060590.1060674},
doi = {10.1145/1060590.1060674},
booktitle = {Proceedings of the Thirty-Seventh Annual ACM Symposium on Theory of Computing},
pages = {563–572},
numpages = {10},
location = {Baltimore, MD, USA},
series = {STOC '05}
}

@misc{flowcutter,
      title={Computing Tree Decompositions with FlowCutter: PACE 2017 Submission}, 
      author={Ben Strasser},
      year={2017},
      eprint={1709.08949},
      archivePrefix={arXiv},
      primaryClass={cs.DS},
      url={https://arxiv.org/abs/1709.08949}, 
}

@ARTICLE{SODA,

  author={Agostini, Nicolas Bohm and Curzel, Serena and Zhang, Jeff Jun and Limaye, Ankur and Tan, Cheng and Amatya, Vinay and Minutoli, Marco and Castellana, Vito Giovanni and Manzano, Joseph and Brooks, David and Wei, Gu-Yeon and Tumeo, Antonino},

  journal={IEEE Micro}, 

  title={Bridging Python to Silicon: The SODA Toolchain}, 

  year={2022},

  volume={42},

  number={5},

  pages={78-88},

  doi={10.1109/MM.2022.3178580}}

@article{liu2017survey,
  title={A survey of deep neural network architectures and their applications},
  author={Liu, Weibo and Wang, Zidong and Liu, Xiaohui and Zeng, Nianyin and Liu, Yurong and Alsaadi, Fuad E},
  journal={Neurocomputing},
  volume={234},
  pages={11--26},
  year={2017},
  publisher={Elsevier}
}

@INPROCEEDINGS{resnet18,

  author={He, Kaiming and Zhang, Xiangyu and Ren, Shaoqing and Sun, Jian},

  booktitle={2016 IEEE Conference on Computer Vision and Pattern Recognition (CVPR)}, 

  title={Deep Residual Learning for Image Recognition}, 

  year={2016},

  volume={},

  number={},

  pages={770-778},

  doi={10.1109/CVPR.2016.90}}

@article{arato_hs,
author = {Arat\'{o}, P\'{e}ter and Mann, Zolt\'{a}n \'{A}d\'{a}m and Orb\'{a}n, Andr\'{a}s},
title = {Algorithmic aspects of hardware/software partitioning},
year = {2005},
issue_date = {January 2005},
publisher = {Association for Computing Machinery},
address = {New York, NY, USA},
volume = {10},
number = {1},
issn = {1084-4309},
url = {https://doi.org/10.1145/1044111.1044119},
doi = {10.1145/1044111.1044119},
journal = {ACM Trans. Des. Autom. Electron. Syst.},
month = jan,
pages = {136–156},
numpages = {21}
}

@inproceedings{critical_path_thesis,
  title={System-level codesign of mixed hardware-software systems},
  author={Asawaree Kalavade and Edward A. Lee},
  year={1995},
  url={https://api.semanticscholar.org/CorpusID:11428750}
}

@book{cygan2015parameterized,
  title={Parameterized algorithms},
  author={Cygan, Marek and Fomin, Fedor V and Kowalik, {\L}ukasz and Lokshtanov, Daniel and Marx, D{\'a}niel and Pilipczuk, Marcin and Pilipczuk, Micha{\l} and Saurabh, Saket},
  volume={5},
  number={4},
  year={2015},
  publisher={Springer}
}

@incollection{althaus2020optimal,
  title={Optimal tree decompositions revisited: A simpler linear-time FPT algorithm},
  author={Althaus, Ernst and Ziegler, Sarah},
  booktitle={Graphs and Combinatorial Optimization: from Theory to Applications: CTW2020 Proceedings},
  pages={67--78},
  year={2020},
  publisher={Springer}
}

@article{xiao2019survey,
  title={A survey on agent-based simulation using hardware accelerators},
  author={Xiao, Jiajian and Andelfinger, Philipp and Eckhoff, David and Cai, Wentong and Knoll, Alois},
  journal={ACM Computing Surveys (CSUR)},
  volume={51},
  number={6},
  pages={1--35},
  year={2019},
  publisher={ACM New York, NY, USA}
}

@inproceedings{sze2017hardware,
  title={Hardware for machine learning: Challenges and opportunities},
  author={Sze, Vivienne and Chen, Yu-Hsin and Emer, Joel and Suleiman, Amr and Zhang, Zhengdong},
  booktitle={2017 IEEE custom integrated circuits conference (CICC)},
  pages={1--8},
  year={2017},
  organization={IEEE}
}

@inproceedings{papadimitriou1988towards,
  title={Towards an architecture-independent analysis of parallel algorithms},
  author={Papadimitriou, Christos and Yannakakis, Mihalis},
  booktitle={Proceedings of the twentieth annual ACM symposium on Theory of computing},
  pages={510--513},
  year={1988}
}

@article{hotpath_dp,
  title={Efficient heuristic algorithms for path-based hardware/software partitioning},
  author={Wu, Jigang and Srikanthan, Thambipillai and Lei, Ting},
  journal={Mathematical and Computer Modelling},
  volume={51},
  number={7-8},
  pages={974--984},
  year={2010},
  publisher={Elsevier}
}

@article{bodlaender1996efficient,
  title={Efficient and constructive algorithms for the pathwidth and treewidth of graphs},
  author={Bodlaender, Hans L and Kloks, Ton},
  journal={Journal of Algorithms},
  volume={21},
  number={2},
  pages={358--402},
  year={1996},
  publisher={Elsevier}
}

@inproceedings{ilp_accurate_high_level,
  title={Accurate high-level modeling and automated hardware/software co-design for effective SoC design space exploration},
  author={Zuo, Wei and Pouchet, Louis-Noel and Ayupov, Andrey and Kim, Taemin and Lin, Chung-Wei and Shiraishi, Shinichi and Chen, Deming},
  booktitle={Proceedings of the 54th Annual Design Automation Conference 2017},
  pages={1--6},
  year={2017}
}

@misc{erde2017directedpathdecompositions,
      title={Directed path-decompositions}, 
      author={Joshua Erde},
      year={2017},
      eprint={1711.00718},
      archivePrefix={arXiv},
      primaryClass={math.CO},
      url={https://arxiv.org/abs/1711.00718}, 
}

@inproceedings{genetic_partitioning,
  title={Functional partitioning for hardware-software codesign using genetic algorithms},
  author={Hidalgo, Jos{\'e} Ignacio and Lanchares, Juan},
  booktitle={EUROMICRO 97. Proceedings of the 23rd EUROMICRO Conference: New Frontiers of Information Technology (Cat. No. 97TB100167)},
  pages={631--638},
  year={1997},
  organization={IEEE}
}

@article{eppstein1992parallel,
  title={Parallel recognition of series-parallel graphs},
  author={Eppstein, David},
  journal={Information and Computation},
  volume={98},
  number={1},
  pages={41--55},
  year={1992},
  publisher={Elsevier}
}

@article{search_survey,
  title={Comparing three heuristic search methods for functional partitioning in hardware--software codesign},
  author={Wiangtong, Theerayod and Cheung, Peter YK and Luk, Wayne},
  journal={Design Automation for Embedded Systems},
  volume={6},
  number={4},
  pages={425--449},
  year={2002},
  publisher={Springer}
}

@article{genetic_algorithm,
  title={Genetic algorithms for hardware--software partitioning and optimal resource allocation},
  author={Purnaprajna, Madhura and Reformat, Marek and Pedrycz, Witold},
  journal={Journal of Systems Architecture},
  volume={53},
  number={7},
  pages={339--354},
  year={2007},
  publisher={Elsevier}
}

@article{system_level,
  title={System level hardware/software partitioning based on simulated annealing and tabu search},
  author={Eles, Petru and Peng, Zebo and Kuchcinski, Krzysztof and Doboli, Alexa},
  journal={Design automation for embedded systems},
  volume={2},
  number={1},
  pages={5--32},
  year={1997},
  publisher={Springer}
}

@article{graph_reduce_hs,
author = {Jiang, Guiyuan and Wu, Jigang and Lam, Siew-Kei and Srikanthan, Thambipillai and Sun, Jizhou},
title = {Algorithmic aspects of graph reduction for hardware/software partitioning},
year = {2015},
issue_date = {June 2015},
publisher = {Kluwer Academic Publishers},
address = {USA},
volume = {71},
number = {6},
issn = {0920-8542},
url = {https://doi.org/10.1007/s11227-015-1381-4},
doi = {10.1007/s11227-015-1381-4},
journal = {J. Supercomput.},
month = jun,
pages = {2251–2274},
numpages = {24}
}

@ARTICLE{integrate_constraints,
  author={Banerjee, Sudarshan and Bozorgzadeh, Elaheh and Dutt, Nikil D.},
  journal={IEEE Transactions on Very Large Scale Integration (VLSI) Systems}, 
  title={Integrating Physical Constraints in HW-SW Partitioning for Architectures With Partial Dynamic Reconfiguration}, 
  year={2006},
  volume={14},
  number={11},
  pages={1189-1202},
  doi={10.1109/TVLSI.2006.886411}}

@article{ants,
author = {Ferrandi, Fabrizio and Lanzi, Pier Luca and Pilato, Christian and Sciuto, Donatella and Tumeo, Antonino},
title = {Ant colony heuristic for mapping and scheduling tasks and communications on heterogeneous embedded systems},
year = {2010},
issue_date = {June 2010},
publisher = {IEEE Press},
volume = {29},
number = {6},
issn = {0278-0070},
url = {https://doi.org/10.1109/TCAD.2010.2048354},
doi = {10.1109/TCAD.2010.2048354},
journal = {Trans. Comp.-Aided Des. Integ. Cir. Sys.},
month = jun,
pages = {911–924},
numpages = {14}
}

@String{Academic = "Academic Press" }

@String{Computer = "{IEEE} Computer" }

@String{Computing = "Computing" }

@STRING{cvpr = {IEEE Proc.~Conf.~on Comp.~Vis.~Patt.~Recog.}}

@STRING{IEEE = {Proc. {IEEE}}}

@String{Springer = "Springer-Verlag" }

@article{staudt2017generating,
  title={Generating realistic scaled complex networks},
  author={Staudt, Christian L and Hamann, Michael and Gutfraind, Alexander and Safro, Ilya and Meyerhenke, Henning},
  journal={Applied Network Science},
  volume={2},
  number={1},
  pages={36},
  year={2017},
  publisher={Springer}
}

@inproceedings{gutfraind2015multiscale,
  title={Multiscale network generation},
  author={Gutfraind, Alexander and Safro, Ilya and Meyers, Lauren Ancel},
  booktitle={Information Fusion (Fusion), 2015 18th International Conference on},
  pages={158--165},
  year={2015},
  organization={IEEE}
}

@article{hager2018multilevel,
  title={A multilevel bilinear programming algorithm for the vertex separator problem},
  author={Hager, William W and Hungerford, James T and Safro, Ilya},
  journal={Computational Optimization and Applications},
  volume={69},
  number={1},
  pages={189--223},
  year={2018},
  publisher={Springer}
}

@INCOLLECTION{buluc2016recent,
  author = {Buluc, Aydin and Meyerhenke, Henning and Safro, Ilya and Sanders,
	Peter and Schulz, Christian},
  title = {Recent advances in graph partitioning},
  booktitle = {Algorithm Engineering: Selected Results and Surveys},
  publisher = {Springer},
  year = {2016},
  volume = {9220},
  pages = {117--158}
}

@inproceedings{liu2022towards,
  title={Towards practical explainability with cluster descriptors},
  author={Liu, Xiaoyuan and Tyagin, Ilya and Ushijima-Mwesigwa, Hayato and Ghosh, Indradeep and Safro, Ilya},
  booktitle={2022 IEEE International Conference on Data Mining Workshops (ICDMW)},
  pages={1--10},
  year={2022},
  organization={IEEE}
}

@inproceedings{liu2022partitioning,
  title={Partitioning dense graphs with hardware accelerators},
  author={Liu, Xiaoyuan and Ushijima-Mwesigwa, Hayato and Ghosh, Indradeep and Safro, Ilya},
  booktitle={International Conference on Computational Science},
  pages={476--483},
  year={2022},
  organization={Springer}
}

@article{liu2022leveraging,
  title={Leveraging special-purpose hardware for local search heuristics},
  author={Liu, Xiaoyuan and Ushijima-Mwesigwa, Hayato and Mandal, Avradip and Upadhyay, Sarvagya and Safro, Ilya and Roy, Arnab},
  journal={Computational Optimization and Applications},
  volume={82},
  number={1},
  pages={1--29},
  year={2022},
  publisher={Springer}
}

\end{document}